\documentstyle[aps,multicol,epsfig,psfig]{revtex}
\newcommand{\bea}{\begin{eqnarray}}
\newcommand{\eea}{\end{eqnarray}}

\begin{document}
\draft
\vfill
\title{Simulation software for 'Simulating Dynamical Features of
  Escape Panic'}

\author{
Dirk Helbing$^{\, 1,2}$,
Ill\'{e}s J. Farkas$^{\, 3}$, 
Tam\'{a}s Vicsek$^{\, 1,3}$
}
\address{
$^{1}$ Collegium Budapest -- Institute for Advanced Study,
  Szenth\'aroms\'ag u. 2, H-1014 Budapest, Hungary \\
$^{2}$ Institute for Economics and Traffic, Dresden University of
  Technology, D-01062 Dresden, Germany\\
$^{3}$ Department of Biological Physics, E\"otv\"os University, P\'azm\'any
  P\'eter S\'et\'any 1A, H-1117 Budapest, Hungary 
}
\maketitle
\thispagestyle{empty}
\begin{abstract}

Simulation software used to produce results in cond-mat/0009448 
-- published as Helbing et.al, Simulating Dynamical Features of Escape
Panic, Nature 407, 487-490 (2000) -- has been made available via
the website of the publication at {\tt http://angel.elte.hu/panic}.

\end{abstract}


\vfill


\section{Brief description}
\label{s_desc}

The source code \cite{panic} 
is in plain text format, and the
files have been successfully compiled and run on various Linux
systems. Documentation is 
available starting from the file {\tt 00\_README.txt} .

\vskip0.2in
The program has three different output formats.
\begin{itemize}
\item The primary output format is through images on the monitor
  (using the so-called X server).
\item The secondary output format is Encapsulated Postscript images.
\item Another possible output format is a compact data file containing
  all coordinates of all particles during the simulation.
\end{itemize}

\vfill

\vskip1truecm
\centerline{\bf Acknowledgements}
\vskip1truecm

The authors thank their host institutions for support, and Z. Csah\'ok
and Z. Farkas for permission to include the two library files they had
previously developed.


\end{document}